\documentclass{article}
\usepackage{graphicx}
\begin{document}

\title{Kaluza-Klein dimensional reduction and Gauss-Codazzi-Ricci equations}
\author{Pei Wang \thanks{peiwang@nwu.edu.cn} \\
Institute of Modern Physics, Northwest University, Xian 710069,
China}
\date{}
\maketitle
\begin{abstract}{In this paper we imitate the traditional method
which is used customarily in the General Relativity and some
mathematical literatures to derive the Gauss-Codazzi-Ricci
equations for dimensional reduction. It would be more distinct
concerning geometric meaning than the vielbein method. Especially,
if the lower dimensional metric is independent of reduced
dimensions the counterpart of the symmetric extrinsic curvature is
proportional to the antisymmetric Kaluza-Klein gauge field
strength. For isometry group of internal space, the SO(n) symmetry
and SU(n) symmetry are discussed. And the Kaluza-Klein instanton
is also enquired.} {PACS:03.70;11.15} {\it Keywords:} Gauss
Codazzi Ricci equation, Kaluza Klein reduction
\end{abstract}

\section{Introduction}
Kaluza-Klein dimensional reduction is a longstanding problem which
is followed with interest by theoretical physicists.\cite{Kaluza}
It is developed from initial unification of gravitational and
electromagnetic interactions to becoming the cornerstone for
superstring and supergravity. (see review papers
\cite{Duff}\cite{Overduin} and references therein.) To depict
Kaluza-Klein dimensional reduction most authors adopt the Cartan
moving frame that is the vielbein method. Its form is elegant and
its algorithm is rapid. But the well-known Gauss-Codazzi-Ricci
equations which describe a submanifold embedded in a Riemann space
are implicit. Perhaps the role played by Codazzi constraint and
Ricci constraint may not clear too. Alternatively we would like to
derive these equations for dimensional reduction by using the
traditional method which is used customarily in the General
Relativity \cite{Hawking}\cite{Wald}\cite{Stephani} and some
mathematic literatures.\cite{Schouten}\cite{Yano} The geometric
meaning may be more distinct than the vielbein method. As a result
we have found the substitute of so called lapse
function\cite{Arnowitt} and shifted function,\cite{Wheeler} that
is, instead of shifted function we have Kaluza Klein gauge
potential and instead of lapse function we have scalar field
tensor. Especially, the symmetric extrinsic curvature tensor now
is replaced by a mixed tensor which has antisymmetric part as well
as symmetric one. When the metric of lower dimensional space is
independent of reduced dimensions the tensor is proportional to
the antisymmetric Kaluza-Klein gauge field strength. \\ The
simplest Kaluza-Klein reduction is through the
torus,\cite{DeWit}\cite{Polchinski} consequently, the isometry
group of internal space is $U(1)^n$. But for connecting with
physics the nonAbelian group is more interest. So we try to
examine SO(n) symmetry and SU(n) symmetry. Up to now there are
only a few of nonlinear ansatz for truncations to the massless
supermultipletes: DeWit and Nicolai demonstrated the consistency
reduced on $S^7$ from 11-dimensional supergravity to the
4-dimensional SO(8) supergravity.\cite{Nicolai} Nastase et al.
found a complete solution for the $S^4$ reduction of
11-dimensional supergravity giving rise to the 7-dimensional
gauged SO(5) supergravity.\cite{Nastase} and a $S^5$ reduction of
IIB supergravity giving rise to the 5-dimensional gauged SO(6)
supergravity was contributed by Cvetic et al.\cite{Cvetic}
Nevertheless as indicated in ref.\cite{Lu} the consistency of
these ansatz always work at the level of the equations of motion.
According to usual understanding, the actions are equivalent to
equations of motion. But these authors (they called above
reduction the Pauli reduction) pointed out that substituting the
ansatz into the higher dimensional action may not give the correct
lower dimensional theory. In this paper we will see the Gauss
equation denotes essentially the reduction relation of action. It
may provide another avenues for further investigation. We know
SU(n) group is a subgroup of SO(2n) there may be byproduct when we
study the spherical reduction, SU(n) ansatz can be embedded in the
SO(2n) ansatz. Besides, in present theory, the Gauss-Codazzi-Ricci
equations are dependent on Kaluza-Klein gauge potential, they may
be defined in different neighborhood (gauge); moreover, we have
yet to study isometric group SU(n), hence except the Kaluza-Klein
monopole we can also enquire the Kaluza-Klein instanton.
\section{Tensor K and vector L,Gauss- Weingarten Formula}
The standard Kaluza Klein reduction formula from D-dimensional
spacetime to d-dimensional subspace is shown in the following
\begin{equation}ds^2=g_{AB}dX^AdX^B\\
=h_{\alpha\beta}dx^\alpha\,dx^\beta+N_{ij}(du^i+N_\alpha^idx^\alpha)(du^j+N_\beta^jdx^\beta)\end{equation}
in which $N_{ij}=N_{ji}$ and $N_\alpha^i$ are the generalization
of lapse function and shifted function in General Relativity
respectively. Physically they represent scalars and gauge fields,
for Abelian theory $N_\alpha^i=A_\alpha^i$, and for nonAbelian
case\cite{Duff}\cite{Overduin}\begin{equation}N_\alpha^i=-\xi_P^iA_\alpha^P,\end{equation}
$\xi_P^i$ are Killing vectors on (D-d)-dimensional internal space
satisfying
\begin{equation} \xi_P^i\partial_i\xi_Q^j-\xi_Q^i\partial_i\xi_P^j=C_{PQ}^R\xi_R^j,\end{equation}
where $C_{PQ}^R$ is the structure constant of isometric group. In
imitation of lapse-shifted method we can introduce the normal
vectors
\begin{equation}
n_A^i=(N_\alpha^i,\delta_j^i),\qquad\,n^{Ai}=(0,N^{-1\,ij}),\qquad\,n^A_in_{Aj}=N^{-1}_{ij}\end{equation}
so that \begin{equation} g_{AB}=h_{AB}+N_{ij}n_A^in_B^j,\qquad
h_{AB}=\left(\begin{array}{cc}h_{\alpha\beta}&0\\0&0\end{array}\right).\end{equation}It
is easy to find the inverse metric
\begin{equation}g^{AB}=h^{AB}+\left(\begin{array}{cc}0&0\\0&N^{-1\,ij}\end{array}\right)\end{equation}where
\begin{equation}h^{AB}=\left(\begin{array}{cc}h^{\alpha\beta}&-N^{\alpha\,j}\\-N^{\beta\,i}&N_\alpha^iN^{\alpha\,i}\end{array}\right).\end{equation}
Obviously, we have
\begin{equation}h_{AB}n^B_i=0=h^{AB}n_B^i.\end{equation} Especially,
there are special components of metric h which satisfy
\begin{equation}
h_\alpha^\beta=\delta_\alpha^\beta,\quad\,h_\alpha^i=-N_\alpha^i,\quad\,h_{iA}=h_{Ai}=h^A_i=0.\end{equation}
We begin by presenting the relation between two
metrices\begin{equation}
h_{\alpha\beta}=h^A_\alpha\,h^B_\beta\,g_{AB}\end{equation} then
we get by differentiating them
\begin{eqnarray*}\lefteqn{D_\gamma\,h_{\alpha\beta}\equiv\partial_\gamma\,h_{\alpha\beta}-N_\gamma^i\partial_ih_{\alpha\beta}}\\
  & &{=(D_\gamma\,h^A_\alpha)h_{A\beta}+(D_\gamma\,h^B_\beta)h_{\alpha\,B}+h^A_\alpha\,h^B_\beta\,h^C_\gamma\partial_Cg_{AB}}\end{eqnarray*}\begin{eqnarray}
  & &{=h^A_\alpha\,h^B_\beta\,h^C_\gamma\partial_Cg_{AB}}\end{eqnarray}
in the last step we have used the equations (9). Define that
\begin{equation}P_{\alpha\beta}^\gamma=\frac{1}{2}h^{\gamma\delta}(D_\alpha\,h_{\delta\beta}+D_\beta\,h_{\alpha\delta}-
D_\delta\,h_{\alpha\beta})\equiv\mathbf{\Gamma}_{\alpha\beta}^\gamma+H_{\alpha\beta}^\gamma,\end{equation}
in which $\mathbf{\Gamma}_{\alpha\beta}^\gamma$ is the Christoffel
symbol in d-dimensional subspace, while\begin{equation}
H_{\alpha\beta}^\gamma\equiv-\frac{1}{2}h^{\gamma\delta}(N_\alpha^i\partial_i\,h_{\delta\beta}+
N_\beta^i\partial_i\,h_{\alpha\delta}-N_\delta^i\partial_i\,h_{\alpha\beta}).\end{equation}
Substituting eq.(11) into eq.(12) we obtain
\begin{eqnarray*}\lefteqn{P_{\alpha\beta}^\gamma=\frac{1}{2}h^{\gamma\delta}h^D_\delta\,h^A_\alpha\,h^B_\beta(\partial_Ag_{DB}+\partial_Bg_{AD}-\partial_Dg_{AB})}\\
  & &{=h^\gamma_Ch^A_\alpha\,h^B_\beta\frac{1}{2}g^{CD}(\partial_Ag_{DB}+\partial_Bg_{AD}-\partial_Dg_{AB})}\end{eqnarray*}\begin{eqnarray}
  & &{=h^\gamma_C(h^A_\alpha\,h^B_\beta\Gamma_{AB}^C+D_\alpha\,h^C_\beta).}\end{eqnarray}
In last equality we have used the ambiguity because of eqs.(9). It
is clear that
\begin{equation}h_C^\gamma\tilde{\nabla}_\alpha\,h_\beta^C
\equiv\,h_C^\gamma(D_\alpha\,h_\beta^C+\Gamma_{AB}^Ch_\alpha^Ah_\beta^B-P_{\alpha\beta}^\delta\,h_\delta^C)=0,\end{equation}
which tells us that $\tilde{\nabla}_\alpha\,h_\beta^C$ is
proportional to the normal vector fields $n^C_i$, hence we can
define tensors $K_{\alpha\beta}^i$ by
\begin{equation}\tilde{\nabla}_\alpha\,h_\beta^C=K^i_{\alpha\beta}n^C_i,\end{equation}
this is the Gauss formula in present case. Operator
$\tilde{\nabla}_\alpha$ we have introduced is an operator which
operates on the D-dimensional index as well as d-dimensional index
simultaneously.\cite{Yano} In fact if we define the following
operators
\begin{equation}
\tilde{\nabla}_\beta\,n^C_i=D_\beta\,n^C_i+\Gamma_{BA}^Ch^B_\beta\,n^A_i,\qquad\tilde{\nabla}_\beta\,n^i_C\\
=D_\beta\,n^i_C-\Gamma_{BC}^Ah^B_\beta\,n^i_A;\end{equation} and
\begin{equation}\tilde{\nabla}_\beta\,u^\gamma=D_\beta\,u^\gamma+P_{\beta\alpha}^\gamma\,u^\alpha,
\qquad\tilde{\nabla}_\beta\,u_\gamma=D_\beta\,u_\gamma-P_{\beta\gamma}^\alpha\,u_\alpha.\end{equation}
Operator $\tilde{\nabla}_\alpha$ on $h^C_\beta$ certainly agrees
with eq.(15). In an earlier version\cite{Wang} we have defined the
operator $\tilde{\nabla}_\alpha$ by using $\partial_\alpha$
instead of $D_\alpha$.As a result, both
$\tilde{\nabla}_\alpha\,g_{AB}$ and
$\tilde{\nabla}_\alpha\,h_{\beta\gamma}$ do not vanish, they
belong to so called nonmetric connection. As a matter of fact in
lower dimensional space the connection is just the torsion free
linear connection described by Schouten \cite{Schouten}; while in
higher dimensional space it is Yano's projective
connection.\cite{Yano} Fortunately, when we redefine operator
$\tilde{\nabla}_\alpha$ as present form the metric property of
connection is recovered. i.e.
\begin{equation}\tilde{\nabla}_\gamma\,g_{AB}=\tilde{\nabla}_\gamma\,h_{\alpha\beta}=0.\end{equation}
From eqs.(14) and (5) we obtain
\begin{eqnarray*}
\lefteqn{\tilde{\nabla}_\alpha\,h^C_\beta=D_\alpha\,h^C_\beta+\Gamma_{AB}^Ch^A_\alpha\,h^B_\beta
     -P_{\alpha\beta}^\gamma\,h^C_\gamma}\\
  & &{=(\delta^C_D-h^\gamma_Dh^C_\gamma)(D_\alpha\,h^D_\beta+\Gamma_{AB}^Dh^A_\alpha\,h^B_\beta)}\end{eqnarray*}\begin{eqnarray}
  & &{=N_{ij}n^j_D(D_\alpha\,h^D_\beta+\Gamma_{AB}^Dh^A_\alpha\,h^B_\beta)n^{Ci}}.\end{eqnarray}Therefore

\begin{equation}
K_{\alpha\beta\,i}\\
  =-\frac{1}{2}[\partial_i\,h_{\alpha\beta}+N_{ij}(D_\alpha\,N^j_\beta
  -D_\beta\,N^j_\alpha)].
  \end{equation}
If there is nonAbelian isometric group on internal manifold, by
means of eqs.(2)and(3) we get
\begin{equation}K_{\alpha\beta\,i}\\
  =-\frac{1}{2}(\partial_i\,h_{\alpha\beta}-N_{ij}\mathcal{F}_{\alpha\beta}^P\xi^j_P),\end{equation}
in which
\begin{equation}\mathcal{F}_{\alpha\beta}^P=\partial_\alpha\mathcal{A}^P_\beta-\partial_\beta\mathcal{A}^P_\alpha
  +C_{QR}^P\mathcal{A}^Q_\alpha\mathcal{A}^R_\beta.\end{equation}
Now the geometric meaning of function $K_{\alpha\beta}^i$ is quite
clear: its symmetric part is a gradient of metric on submanifold
which vanishes if neglect massive particles in the compactified
theory. The antisymmetric part proportional to a Yang-Mills gauge
field, which really can be thought of a kind of "curvature". We
know in old Gauss-Codazzi-Ricci theory the corresponding K tensor
is a symmetric extrinsic curvature.\\
Next, we write down the Weingarten formula
\begin{eqnarray*}
\lefteqn{\tilde{\nabla}_\beta\,n^i_A=h^\alpha_Ah^B_\alpha\tilde{\nabla}_\beta\,n^i_B+N_{kl}n^k_An^{Bl}\tilde{\nabla}_\beta\,n^i_B}\\
  & &{=-h^\alpha_An^i_B\tilde{\nabla}_\beta\,h^B_\alpha+n^k_AL_{\beta\,k}{}^i}
  \end{eqnarray*}
  \begin{eqnarray}
  & &{=-h^\alpha_A\tilde{K}_{\beta\alpha}{}^i+n^k_AL_{\beta\alpha}{}^i,}
  \end{eqnarray}
in which
\begin{equation}\tilde{K}_{\beta\alpha}^i=K_{\beta\alpha}^jN^{-1\,i}_j,\end{equation}
\begin{equation}L_{\beta\,ji}\equiv\,N_{jk}n^{Ak}\tilde{\nabla}_\beta\,n_{Ai}
=-\frac{1}{2}N^{-1\,l}_i(D_\beta\,N_{jl}+N_{lk}\partial_jN^k_\beta-N_{jk}\partial_lN^k_\beta),\end{equation}
or\begin{equation}\tilde{L}_{\beta\,ij}=N^{-1\,k}_iL_{\beta\,kj}\\
=\frac{1}{2}(D_\beta\,N^{-1}_{ij}-N^{-1\,l}_i\partial_lN_{\beta\,j}+N^{-1\,l}_j\partial_lN_{\beta\,i}),\end{equation}
which satisfies
\begin{equation}\tilde{L}_{\beta\,ji}+\tilde{L}_{\beta\,ij}=D_\beta\,N^{-1}_{ij},\end{equation}
and for nonAbelian case
\begin{equation}\tilde{L}_{\beta\,ij}=\frac{1}{2}D_\beta\,N^{-1}_{ij}-(N^{-1\,l}_i\partial_l\xi_{Pj}
-N^{-1\,l}_j\partial_l\xi_{Pi})\mathcal{A}^P_\beta.\end{equation}
\section{ Gauss-Codazzi-Ricci Equations}
From definition of $\tilde{\nabla}_\alpha\,h^C_\beta$ it is
straightforward to calculate the antisymmetric double derivative
of $h^C_\beta$ as follows
\begin{eqnarray*}\lefteqn{h^\delta_C(\tilde{\nabla}_\gamma\tilde{\nabla}_\alpha-\tilde{\nabla}_\alpha\tilde{\nabla}_\gamma)h^C_\beta}\\
  & &{=h^\delta_Ch^A_\alpha\,h^B_\beta\,h^D_\gamma\,{R_{ADB}}^C-{S_{\alpha\gamma\beta}}^\delta}\end{eqnarray*}
  \begin{eqnarray}
  & &{+h^{\delta\eta}{N^{-1}}^{ij}(\partial_i\,h_{\alpha\gamma}+2K_{\alpha\gamma\,i})K_{\beta\eta\,j}}
  \end{eqnarray}
in which ${R_{ADB}}^C$ is the Riemann curvature tensor of higher
dimensional space and
\begin{eqnarray*}
\lefteqn{{S_{\alpha\gamma\beta}}^\delta\equiv\,D_\gamma\,P_{\alpha\beta}^\delta-D_\alpha\,P_{\gamma\beta}^\delta
     +P_{\gamma\eta}^\delta\,P_{\alpha\beta}^\eta-P_{\alpha\eta}^\delta\,P_{\gamma\beta}^\eta}\\
  & &{={\mathbf{R}_{\alpha\gamma\beta}}^\delta+N^i_\alpha\,\partial_i\mathbf{\Gamma}_{\gamma\beta}^\delta-N^i_\gamma\partial_i\mathbf{\Gamma}_{\alpha\beta}^\delta
     +D_\gamma\,H_{\alpha\beta}^\delta-D_\alpha\,H_{\gamma\beta}^\delta}\\
  & &{+H_{\gamma\eta}^\delta\,H_{\alpha\beta}^\eta-H_{\alpha\eta}^\delta\,H_{\gamma\beta}^\eta
     +\mathbf{\Gamma}_{\gamma\eta}^\delta\,H_{\alpha\beta}^\eta+\mathbf{\Gamma}_{\alpha\beta}^\eta\,H_{\gamma\eta}^\delta}
     \end{eqnarray*}
     \begin{eqnarray}
  & &{-\mathbf{\Gamma}_{\alpha\eta}^\delta\,H_{\gamma\beta}^\eta-\mathbf{\Gamma}_{\gamma\beta}^\eta\,H_{\alpha\eta}^\delta,}
  \end{eqnarray}

where
\begin{equation}
{\mathbf{R}_{\alpha\gamma\beta}}^\delta=\partial_\gamma\mathbf{\Gamma}_{\alpha\beta}^\delta
  -\partial_\alpha\mathbf{\Gamma}_{\gamma\beta}^\delta+\mathbf{\Gamma}_{\gamma\zeta}^\delta\mathbf{\Gamma}_{\alpha\beta}^\zeta
  -\mathbf{\Gamma}_{\alpha\zeta}^\delta\mathbf{\Gamma}_{\gamma\beta}^\zeta\end{equation}
is the Riemann curvature tensor in lower dimensional space. On the
other hand
\begin{eqnarray*}\lefteqn{h^\delta_C(\tilde{\nabla}_\gamma\tilde{\nabla}_\alpha
     -\tilde{\nabla}_\alpha\tilde{\nabla}_\gamma)h^C_\beta}\\
  & &{=\tilde{\nabla}_\alpha\,h^\delta_C\tilde{\nabla}_\gamma\,h^C_\beta
     -\tilde{\nabla}_\gamma\,h^\delta_C\tilde{\nabla}_\alpha\,h^C_\beta}\end{eqnarray*}\begin{eqnarray}
  & &{=N^{-1}_{ij}h^{\delta\eta}(K_{\alpha\eta}^iK_{\gamma\beta}^j-K_{\gamma\eta}^iK_{\alpha\beta}^j).}\end{eqnarray}
Equating right hand sides of (30) and (33) and contracting
$\gamma$ with $\delta$ and $\alpha$ with $\beta$ we obtain the
Gauss equation
\begin{eqnarray*}\lefteqn{h^{AB}h^{CD}R_{ADBC}=h^{\alpha\beta}{S_{\alpha\gamma\beta}}^\gamma}\\
  & &{+h^{\alpha\beta}h^{\gamma\delta}N^{-1}_{ij}(K_{\alpha\delta}^iK_{\gamma\beta}^j
     -K_{\gamma\delta}^iK_{\alpha\beta}^j-2K_{\alpha\gamma}^iK_{\beta\delta}^j)}\end{eqnarray*}\begin{eqnarray}
  & &{+\frac{1}{2}h^{\alpha\beta}h^{\gamma\delta}N^{-1\,ij}\partial_ih_{\alpha\gamma}\partial_jh_{\beta\delta}.}\end{eqnarray}
Because of
\begin{eqnarray*}\lefteqn{h^{AB}h^{CD}R_{ADBC}=R-2N^{-1\,ij}R_{ij}+N^{-1\,ij}N^{-1\,kl}R_{ikjl}}\\
  & &=R-h^{\alpha\beta}h^{\gamma\delta}N^{-1\,ij}[2K_{\alpha\gamma\,i}K_{\beta\delta\,j}-(N_{ik}\partial_jN^k_\delta\\
  & &-\frac{1}{2}D_\delta\,N_{ij})(N^k_\gamma\partial_kh_{\alpha\beta}
     -N^k_\alpha\partial_kh_{\beta\gamma}-N^k_\beta\partial_kh_{\alpha\gamma})]\end{eqnarray*}\begin{eqnarray}
  & &{-X(N^i_\alpha,N_{kl}),}\end{eqnarray}
\begin{eqnarray*}\lefteqn{X(N^i_\alpha,N_{kl})\equiv}\\
  & &N^{-1\,ij}\{h^{\alpha\beta}[2D_\alpha\,N^k_\beta\partial_iN_{jk}+2D_\alpha(N_{ik}\partial_jN^k_\beta)-\partial_i\partial_jh_{\alpha\beta}\\
  & &-D_\alpha\,D_\beta\,N_{ij}-D_\alpha\,N^k_\beta\partial_kN_{ij}-2N_{kl}\partial_iN^k_\alpha\partial_jN^l_\beta]\\
  & &+\frac{1}{2}h^{\alpha\beta}N^{-1\,kl}[N_{im}\partial_lN^m_\alpha(3N_{jn}\partial_kN^n_\beta\\
  & &-N_{kn}\partial_jN^n_\beta)-2N_{im}\partial_jN^m_\alpha\,N_{ln}\partial_kN^n_\beta\\
  & &-2N_{im}(\partial_lN^m_\alpha\,D_\beta\,N_{jk}-\partial_jN^m_\alpha\,D_\beta\,N_{lk})\\
  & &+\frac{1}{2}(3D_\alpha\,N_{ik}D_\beta\,N_{jl}-D_\alpha\,N_{ij}D_\beta\,N_{kl})-(2\partial_iN_{jk}\\
  & &-\partial_kN_{ij})(2N_{lm}D_\beta\,N^m_\alpha-\partial_lh_{\alpha\beta})]\\
  & &+N^{-1\,kl}(\partial_i\partial_kN_{jl}-\partial_i\partial_kN_{kl})+\frac{1}{4}N^{-1\,kl}N^{-1\,mn}[2\partial_iN_{mk}(\partial_lN_{jn}\\
  & &+\partial_jN_{nl}-2\partial_nN_{jl})+\partial_mN_{ik}\partial_nN_{jl}-\partial_mN_{ij}\partial_nN_{kl}\\
  & &-4\partial_iN_{jm}(\partial_lN_{kn}-\partial_nN_{kl})]\};\end{eqnarray*}\begin{equation}{}\end{equation}
and\begin{equation}h^{\alpha\beta}{S_{\alpha\gamma\beta}}^\gamma=\mathbf{R}+V(h_{\alpha\beta}),\end{equation}
\begin{eqnarray*}\lefteqn{V(h_{\alpha\beta})\equiv\,N^i_\alpha\partial^\gamma\,h^{\alpha\beta}\partial_ih_{\beta\gamma}}\\
  & &{+h^{\alpha\beta}(\partial^\gamma\,N^i_\gamma\partial_ih_{\alpha\beta}-\partial^\gamma\,N^i_\alpha\partial_ih_{\beta\gamma})}\\
  & &+\frac{1}{2}N^{\alpha\,i}[\partial_\alpha(h^{\gamma\delta}\partial_ih_{\gamma\delta})\\
  & &+h^{\gamma\delta}\partial_\alpha\partial_ih_{\gamma\delta}]+\frac{1}{4}h^{\alpha\beta}(2N^{\gamma\,i}N^{\delta\,j}\\
  & &-h^{\gamma\delta}N^i_\eta\,N^{\eta\,j})(\partial_ih_{\alpha\beta}\partial_jh_{\gamma\delta}\\
  & &-\partial_ih_{\alpha\delta}\partial_jh_{\beta\gamma})
     +h^{\alpha\beta}h^{\gamma\delta}(N^i_\gamma\partial_iN^j_\alpha\partial_jh_{\delta\beta}\\
  & &-N^i_\gamma\partial_iN^j_\delta\partial_jh_{\alpha\beta}+N^i_\gamma\,N^j_\alpha\partial_i\partial_jh_{\delta\beta}\\
  & &-N^i_\gamma\,N^j_\delta\partial_i\partial_jh_{\alpha\beta})\\
  & &+\frac{1}{2}h^{\alpha\beta}(2N^i_\gamma\,N^j_\alpha\partial_jh_{\beta\delta}\end{eqnarray*}\begin{eqnarray}
  & &-N^i_\alpha\,N^j_\beta\partial_jh_{\gamma\delta}
     -N^i_\gamma\,N^j_\delta\partial_jh_{\alpha\beta})\partial_ih^{\gamma\delta},\end{eqnarray}
thus
\begin{eqnarray*}\lefteqn{R=\mathbf{R}}\\
  & &+h^{\alpha\beta}h^{\gamma\delta}N^{-1\,ij}[K_{\alpha\delta\,i}K_{\gamma\beta\,j}
     -K_{\gamma\delta\,i}K_{\alpha\beta\,j}\\
  & &+\frac{1}{2}\partial_ih_{\alpha\gamma}\partial_jh_{\beta\delta}-(N_{ik}\partial_jN^k_\delta
     -\frac{1}{2}D_\delta\,N_{ij})(N^k_\gamma\partial_kh_{\alpha\beta}\\
  & &-N^k_\alpha\partial_kh_{\beta\gamma}-N^k_\beta\partial_kh_{\alpha\gamma})]\end{eqnarray*}\begin{eqnarray}
  & &+X(N^i_\alpha,N_{kl})+V(h_{\alpha\beta}).\end{eqnarray}
It is well-known that the Lagrangian of Einstein gravitational
equation is $\sqrt{-g}R$, hence, we have to do a conformal
transformation first to assure that Gauss equation represents the
Kaluza-Klein reduction of action from D-dimensional gravitation to
d-dimensional gravitation plus matters (gauge fields and scalars).
Since
\begin{equation}g_{AB}=\left(\begin{array}{cc}h_{\alpha\beta}+N_{ij}N^i_\alpha\,N^j_\beta&N_{ij}N^i_\alpha\\N_{ij}N^j_\beta&N_{ij}\end{array}\right)\end{equation}
from the triangularization of its vielbein form it is easy to
realize
\begin{equation}det\,g_{AB}=det\,h_{\alpha\beta}det\,N_{ij}.\end{equation}
Therefore we adopt the following conformal transformation
\begin{equation}g_{AB}\rightarrow\hat{g}_{AB}=(det\,N_{ij})^{-\frac{1}{D-2}}g_{AB}.\end{equation}
We then obtain
\begin{eqnarray*}\lefteqn{\sqrt{-\hat{g}}\hat{R}=\sqrt{-h}[R+U(det\,N)]}\\
  & &=\sqrt{-h}\{\mathbf{R}+h^{\alpha\beta}h^{\gamma\delta}N^{-1\,ij}[K_{\alpha\delta\,i}K_{\gamma\beta\,j}
     -K_{\gamma\delta\,i}K_{\alpha\beta\,j}\\
  & &+\frac{1}{2}\partial_ih_{\alpha\gamma}\partial_jh_{\beta\delta}-(N_{ik}\partial_jN^k_\delta
     -\frac{1}{2}D_\delta\,N_{ij})(N^k_\gamma\partial_kh_{\alpha\beta}\\
  & &-N^k_\alpha\partial_kh_{\beta\gamma}-N^k_\beta\partial_kh_{\alpha\gamma})]
     +X(N^i_\alpha,N_{kl})+V(h_{\alpha\beta})+U(det\,N)\},\end{eqnarray*}\begin{equation}{}\end{equation}
\begin{eqnarray*}\lefteqn{U(det\,N)\equiv}\\
  & &\frac{D-1}{D-2}\{D^\alpha\,D_\alpha\,ln\,det\,N-\frac{1}{4}D^\alpha\,ln\,det\,N\,D_\alpha\,ln\,det\,N
     +N^{-1\,ij}[\partial_i\partial_jln\,det\,N\\
  & &-\frac{1}{4}\partial_iln\,det\,N\partial_jln\,det\,N+\frac{1}{2}h^{\alpha\beta}\partial_ih_{\alpha\beta}\partial_jln\,det\,N\\
  & &-N^{-1\,kl}(\partial_kN_{il}-\frac{1}{2}\partial_iN_{kl})\partial_jln\,det\,N]\\
  & &+\frac{1}{2}h^{\gamma\delta}[h^{\alpha\beta}(N^j_\alpha\partial_jh_{\delta\beta}
     +N^j_\beta\partial_jh_{\delta\alpha}\end{eqnarray*}\begin{eqnarray}
  & &-N^j_\delta\partial_jh_{\alpha\beta})-2\partial_kN^k_\delta+N^{-1\,kl}D_\delta\,N_{kl}]D_\gamma\,ln\,det\,N\}.\end{eqnarray}
Equation (43) gives the Lagrangian reduction formula.\\ Next, we
would like to find the Codazzi equation and Ricci equation.
Following definition (17) we have
\begin{eqnarray*}\lefteqn{(\tilde{\nabla}_\gamma\tilde{\nabla}_\beta-\tilde{\nabla}_\beta\tilde{\nabla}_\gamma)n^i_C}\\
  & &={R_{DBC}}^Ah^B_\beta\,h^D_\gamma\,n^i_A-\frac{1}{2}N^{-1\,ik}(D_\beta\,N^j_\gamma
     -D_\gamma\,N^j_\beta)[(D_\alpha\,N_{jk}\\
  & &-N_{kl}\partial_jN^l_\alpha-N_{jl}\partial_kN^l_\alpha)h^\alpha_C
     +(\partial_lN_{jk}+\partial_jN_{lk}-\partial_kN_{jl})n^l_C]\end{eqnarray*}\begin{equation}{}\end{equation}
on the one hand, and through the Weingarten formula (24) we know
\begin{eqnarray*}\lefteqn{(\tilde{\nabla}_\gamma\tilde{\nabla}_\beta-\tilde{\nabla}_\beta\tilde{\nabla}_\gamma)n^i_C}\\
  & &{=-(\tilde{\nabla}_\gamma\tilde{K}_{\beta\alpha}^i-\tilde{\nabla}_\beta\tilde{K}_{\gamma\alpha}^i)h^\alpha_C
     -(\tilde{K}_{\beta\alpha}^i{K_\gamma}^\alpha_j-\tilde{K}_{\gamma\alpha}^i{K_\beta}^\alpha_j)n^j_C}\\
  & &{+(\tilde{\nabla}_\gamma{L_{\beta\,j}}^i
     -\tilde{\nabla}_\beta{L_{\gamma\,j}}^i)n^j_C-(\tilde{K}_{\gamma\alpha}^j{L_{\beta\,j}}^i
     -\tilde{K}_{\beta\alpha}^j{L_{\gamma\,j}}^i)h^\alpha_C}\end{eqnarray*}\begin{eqnarray}
  & &{+({L_{\gamma\,k}}^j{L_{\beta\,j}}^i-{L_{\beta\,k}}^j{L_{\gamma\,j}}^i)n^k_C}\end{eqnarray}
on the other hand. Comparing both expressions we obtain the two
equations immediately. The Codazzi equation is
\begin{eqnarray*}\lefteqn{\tilde{\nabla}_\beta\tilde{K}_{\gamma\alpha}^i-\tilde{\nabla}_\gamma\tilde{K}_{\beta\alpha}^i
     +\tilde{K}_{\beta\alpha}^j{L_{\gamma\,j}}^i-\tilde{K}_{\gamma\alpha}^j{L_{\beta\,j}}^i}\\
  & &{-N^{-1\,ik}N^{-1\,jl}(D_\alpha\,N_{jk}-N_{km}\partial_jN^m_\alpha
     -N_{jm}\partial_kN^m_\alpha)(K_{\beta\gamma\,l}+\frac{1}{2}\partial_lh_{\beta\gamma})}\end{eqnarray*}\begin{eqnarray}
  & &{={R_{DBC}}^Ah^B_\beta\,h^D_\gamma\,h^C_\alpha\,n^i_A.}\end{eqnarray}
After contracting index $\alpha$ with $\gamma$ we get the
following form
\begin{eqnarray*}\lefteqn{\tilde{\nabla}_\beta{\tilde{K}_\alpha}^{\alpha\,i}
     -\tilde{\nabla}_\alpha{\tilde{K}_\beta}^{\alpha\,i}}\\
  & &{+{\tilde{K}_\beta}^{\alpha\,j}{L_{\alpha\,j}}^i-{\tilde{K}_\alpha}^{\alpha\,j}{L_{\beta\,j}}^i}\\
  & &-\frac{1}{2}N^{-1\,ki}N^{-1\,lj}\{D_\beta\partial_kN_{jl}-D_\beta\partial_lN_{kj}
     +\partial_jD_\beta\,N_{kl}+(\partial_lN_{km}\\
  & &-\partial_kN_{lm})\partial_jN^m_\beta+2\partial_j(N^m_lL_{\beta\,km})
     -2h^{\gamma\alpha}[K_{\beta\gamma\,k}(N^m_lL_{\alpha\,jm}\\
  & &+N_{lm}\partial_jN^m_\alpha)
     +(K_{\beta\gamma\,l}+\partial_lh_{\beta\gamma})(N^m_kL_{\alpha\,jm}+N_{km}\partial_jN^m_\alpha)]\\
  & &-{L_{\beta\,k}}^m(\partial_jN_{lm}+\partial_lN_{jm}-\partial_mN_{jl})+{L_{\beta\,l}}^m(\partial_kN_{jm}
     +\partial_jN_{km}-\partial_mN_{jk})\}\end{eqnarray*}\begin{eqnarray}
  & &{=R_{BA}h^B_\beta\,n^{Ai}.}\end{eqnarray}
To keep consistency with the Lagrangian reduction we also perform
conformal transformation (42) for Codazzi equation
\begin{equation}\hat{R}_{BA}h^B_\beta\,n^{Ai}=R_{BA}h^B_\beta\,n^{Ai}+W(det\,N)=\mathcal{T}_{BA}h^B_\beta\,n^{Ai},\end{equation}
\begin{eqnarray*}\lefteqn{W(det\,N)}\\
  & &\equiv\frac{1}{2}N^{-1\,ij}[D_\beta\partial_j\,ln\,det\,N
     +\frac{1}{2(D-2)}D_\beta\,ln\,det\,N\partial_j\,ln\,det\,N\\
  & &+N^{-1\,ij}h^{\gamma\delta}K_{\beta\delta\,j}D_\gamma\,ln\,det\,N
     -\frac{1}{2}N^{-1\,kl}(D_\beta\,N_{lj}+N_{lm}\partial_jN^m_\beta\\
  & &-N_{jm}\partial_lN^m_\beta)\partial_k\,ln\,det\,N].\end{eqnarray*}\begin{equation}{}\end{equation}
In the last step we have used the higher dimensional Einstein
equation where matter field tensor $\mathcal{T}_{BA}$ is not equal
to zero if D-dimensional spacetime is not pure gravity. In short
eq.(49) gives out a constraint.\\
Finally, we write down the Ricci equation as following
\begin{eqnarray*}\lefteqn{\tilde{\nabla}_\gamma{L_{\beta\,j}}^i-\tilde{\nabla}_\beta{L_{\gamma\,j}}^i
     +\tilde{K}_{\gamma\alpha}^i{K_\beta}^\alpha_j-\tilde{K}_{\beta\alpha}^i{K_\gamma}^\alpha_j}\\
  & &{+{L_{\gamma\,j}}^k{L_{\beta\,k}}^i-{L_{\beta\,j}}^k{L_{\gamma\,k}}^i
     -N^{-1\,ik}N^{-1\,lm}(\partial_lN_{jk}+\partial_jN_{lk}-\partial_kN_{jl})(K_{\beta\gamma\,m}
     +\frac{1}{2}\partial_mh_{\beta\gamma})}\end{eqnarray*}\begin{eqnarray}
  & &{={R_{DBC}}^Ah^B_\beta\,h^D_\gamma\,N^k_jn^i_An^C_k=N^{-1\,ik}R_{DBjk}h^B_\beta\,h^D_\gamma.}\end{eqnarray}
A simplest constraint occurs if i=j, i.e. the right hand side of
eq.(51) vanishes. Other constraint works when above Ricci equation
is combined with Gauss equation.
\section{Some examples}
\subsection{Isometric group SO(n) and SU(n)}
Let us consider the isometric group SO(n) first. Thus, a spherical
internal space $S^{n-1}(n=D-d+1)$is suitable for present topic,
and the spherical harmonics will be good
instrument.\cite{Salam}\cite{Nastase} The Killing vector can be
written as
\begin{equation}V^{IJ}_i=y^{[I}\partial_iy^{J]}\equiv\frac{1}{2}(y^I\partial_iy^J-y^J\partial_iy^I).\end{equation}
By using properties of spherical
harmonics\begin{equation}y^Iy_I=1,\qquad
\partial_iy^I\partial^iy^J+y^Iy^J=\delta^{IJ}\end{equation}and so
on, we can show that
\begin{equation}\partial_iV^{IJ}_j+\partial_jV^{IJ}_i=0,\end{equation}
and
\begin{equation}V^{IJ}_i\partial^iV^{KL}_j-V^{KL}_i\partial^iV^{IJ}_j=\frac{1}{2}(\delta^{JK}V^{IL}_j-\delta^{JL}V^{IK}_j
+\delta^{IL}V^{JK}_j-\delta^{IK}V^{JL}_j),\end{equation}which is
the commutator of so(n) algebra. In this prescription
\begin{equation}N^i_\alpha=-V^i_{IJ}(u)\mathcal{A}^{IJ}_\alpha(x)\equiv\,N^I_\alpha\partial^iy_I\end{equation}
\begin{equation}N^I_\alpha\equiv\mathcal{A}^{IJ}_\alpha\,y_J=(\mathbf{L}\cdot\mathcal{A}_\alpha)^{IJ}y_J\end{equation}
in which $\mathbf{L}$ is the generator of so(n).\\
If we introduce the scalar field tensor $T_{IJ}(x)$ which was used
by authors of ref.\cite{Nastase} then
\begin{equation}N_{ij}=\Delta^{-1}T^{-1}_{IJ}\partial_iy^I\partial_jy^J,\qquad
\Delta\equiv\,T_{IJ}y^Iy^J,\end{equation} and \begin{equation}
N^{-1}_{ij}=2V^{IK}_iV^{JL}_jT_{IJ}T_{KL}.\end{equation} Now we
find
\begin{equation}K_{\alpha\beta\,i}=-\frac{1}{2}[\partial_ih_{\alpha\beta}
+\Delta^{-1}T^{-1}_{IJ}y_K\partial_iy^I(\mathbf{L}\cdot\mathcal{F})^{KJ}].\end{equation}
(in later ansatz metric $h_{\alpha\beta}$ is supposed independent
of u) And the metric becomes
\begin{eqnarray*}\lefteqn{ds^2=h_{\alpha\beta}dx^\alpha\,dx^\beta}\\
  & &{+\Delta^{-1}T^{-1}_{IJ}[dy^I+(\mathbf{L}\cdot\mathcal{A}_\alpha)^{IK}y_Kdx^\alpha][dy^J
     +(\mathbf{L}\cdot\mathcal{A}_\beta)^{JL}y_Ldx^\beta]}\end{eqnarray*}\begin{equation}{}\end{equation}
In form, it looks like a D+1 dimensional metric, but with a
constraint $y^Iy_I=1$.\\
In fact, by using of these gauge fields and scalar fields the
authors of ref.\cite{Nastase} found a full nonlinear ansatz
truncated to massless fields for 11-dimensional supergravity
reduced to 7-dimensional spacetime through $S^4$ spherical
reduction in which a form field ansatz $F_{(4)}$ guaranteed the
consistency. But there is a Chern-Simons FFA term which makes
things a little complicated. Follow closely another group of
authors found a full nonlinear ansatz for 10-dimensional IIB
supergravity reduced to 5 spacetime on $S^5$ which is particularly
relevant for AdS/CFT correspondence.\cite{Cvetic} In this ansatz
except the 10-dimensional gravitation we have a selfdual 5 form
field. By means of this example we may use D=10 to d=5 Gauss
equation to reduce the system
\begin{equation}\mathcal{L}=\sqrt{-\hat{g}}(\hat{R}_{(10)}-\frac{1}{5!}G_{ABCDE}G^{ABCDE})\end{equation}
in which ansatz $G_{(5)}$ is given in \cite{Cvetic} (let coupling
constant g=1)
\begin{eqnarray*}\lefteqn{G_{\alpha\beta\gamma\delta\epsilon}=-U{\epsilon}_{\alpha\beta\gamma\delta\epsilon}}\\
  & &{+T^{-1}_{IJ}\epsilon_{\alpha\beta\gamma\delta\eta}\mathcal{D}^\eta\,T^{JK}y_K(\mathbf{L}\cdot\mathcal{A}_\epsilon)^{IL}y_L}\\
  & &-\frac{1}{2}T^{-1}_{IK}T^{-1}_{JL}\epsilon_{\alpha\beta\gamma\eta\zeta}(\mathbf{L}\cdot\mathcal{F}^{\eta\zeta})^{IJ}[(\mathbf{L}\cdot\mathcal{A}_\delta)^{KM}y_M(\mathbf{L}\cdot\mathcal{A}_\epsilon)^{LN}y_N\\
  & &-(\delta\leftrightarrow\epsilon)]\end{eqnarray*}\begin{equation}{}\end{equation}
\begin{eqnarray*}\lefteqn{G_{\alpha\beta\gamma\delta\,i}}\\
  & &=T^{-1}_{IJ}{\epsilon}_{\alpha\beta\gamma\delta\eta}\mathcal{D}^\eta\,T^{JK}y_K\partial_iy^I
      -T^{-1}_{IK}T^{-1}_{JL}\epsilon_{\alpha\beta\gamma\eta\zeta}(\mathbf{L}\cdot\mathcal{F}^{\eta\zeta})^{IJ}(\mathbf{L}\cdot\mathcal{A}_\delta)^{KM}y_M\partial_iy^L
\end{eqnarray*}\begin{equation}{}\end{equation}
\begin{equation}G_{\alpha\beta\gamma\,ij}=-T^{-1}_{IK}T^{-1}_{JL}\epsilon_{\alpha\beta\gamma\eta\zeta}(\mathbf{L}\cdot\mathcal{F}^{\eta\zeta})^{IJ}\partial_iy^K\partial_jy^L,\end{equation}where
\begin{equation}U\equiv2T_{IJ}T^{JK}y^Iy_K-\Delta\,T^I_I,\qquad\Delta\equiv\,T_{IJ}y^Iy^J,\end{equation}
\begin{equation}(\mathbf{L}\cdot\mathcal{F})^{IJ}=d(\mathbf{L}\cdot\mathcal{A})^{IJ}+(\mathbf{L}\cdot\mathcal{A})^{IK}\wedge(\mathbf{L}\cdot\mathcal{A})_K^J,\end{equation}
\begin{equation}\mathcal{D}_\alpha\,T_{IJ}=\partial_\alpha\,T_{IJ}+(\mathbf{L}\cdot\mathcal{A}_\alpha)_{IK}T^K_J+T_{IK}(\mathbf{L}\cdot\mathcal{A}_\alpha)_J^K.\end{equation}
Following the ordinary logic, to substitute 10-dimensional
formula(43) into (62) we ought to gain the 5-dimensional
Lagrangian, i.e. the $\mathcal{L}_5$ in ref.\cite{Cvetic}.
However, gazing at eq.(43) it seems difficult to get the expectant
result. In view of demonstration of consistency of known ansatz is
only at the level of equations of motion, it needs more effort for
checking ansatz with action. Because the calculation is
complicated, so we prefer to let them for further investigation.\\
Moreover, we need energy-momentum tensor for Codazzi constrain
\begin{equation}\mathcal{T}_{AB}=-\frac{1}{4!}(G_{ACDEF}{G_B}^{CDEF}-\frac{1}{10}\hat{g}_{AB}G_{ABCDE}G^{ABCDE}).\end{equation}\\
Next,we examine isometric group SU(n). Since SU(n) is a subgroup
of SO(2n), we may still use the spherical harmonics to describe
metric and others. Let$\mathbf{T}$ be the SU(n) generator in
2n-dimensional representation, and $\mathbf{t}$ in basic
representation. To characterize Kaluza-Klein reduction for
isometric group SU(n) what we have to do is to change SO(n)
generator $\mathbf{L}$ to SU(n) generator $\mathbf{T}$. Especially
we now have
\begin{equation}N^I_\alpha=(\mathbf{T}\cdot\mathcal{A})^{IJ}y_J.\end{equation}
Because the number of spherical harmonics is even, we may arrange
them in pair. Let $a=i,\cdots,n$ be the first half of I, we choose
that \begin{equation}z^a=y^a+iy^{a+n},\end{equation} and
\begin{equation}(\mathbf{t})^{ab}=(\mathbf{T})^{ab}+i(\mathbf{T})^{a+n\,b},\end{equation}
where \begin{equation}(\mathbf{T})^{ab}=(\mathbf{T})^{a+n\,b+n},
\qquad (\mathbf{T})^{a\,b+n}=-(\mathbf{T})^{a+n\,b},\end{equation}
so that
\begin{eqnarray}\lefteqn{N^a_\alpha\equiv\,N^{I=a}_\alpha+iN^{I=a+n}\equiv{}^{\Re}N^a_\alpha+i{}^{\Im}N^a_\alpha}\\
  & &{=(\mathbf{T}\cdot\mathcal{A})^{aI}y_I+i(\mathbf{T}\cdot\mathcal{A})^{a+n\,I}y_I=(\mathbf{t}\cdot\mathcal{A})^{ab}z_b.}\end{eqnarray}
Two special examples are (i) SU(2)
\begin{equation}\mathbf{t}=\frac{1}{2i}\mathbf{\tau}, \qquad
\mathbf{\tau}\sim\,Pauli\quad matrix,
\end{equation}\begin{equation}\mathbf{T}=\frac{1}{2}\mathbf{\Sigma},\quad
\Sigma_1=L_{14}+L_{23},\quad \Sigma_2=-(L_{34}+L_{12}), \quad
\Sigma_3=-(L_{24}+L_{31}); \end{equation}(ii) SU(3)
\begin{equation}\mathbf{t}=\frac{1}{2i}\mathbf{\lambda}, \qquad
\mathbf{\lambda}\sim\,Gell-Mann\quad matrix,\end{equation}
\begin{equation}\mathbf{T}=\frac{1}{2}\mathbf{\Lambda},\quad\Lambda_1=L_{15}+L_{24},\quad\Lambda_2=-(L_{12}+L_{45}),
\quad\Lambda_3=L_{14}-L_{25},\end{equation}\begin{equation}\Lambda_4=L_{16}+L_{34},\quad\Lambda_5=-(L_{13}+L_{46}),
\quad\Lambda_6=L_{26}+L_{35},\end{equation}\begin{equation}
\quad\Lambda_7=-(L_{23}+L_{56}),\quad\Lambda_8=\frac{1}{\sqrt{3}}(L_{14}+L_{25}-2L_{36}).\end{equation}
Suppose that the scalar tensor $T_{IJ}$ keeps in real and
possesses block diagonal form. The subset of "SO(2n) metric" (61)
becomes \begin{equation}
g_{AB}=\left(\begin{array}{ccc}h_{\alpha\beta}+\Delta^{-1}T^{-1}_{ab}[{}^{\Re}N^a_\alpha{}^{\Re}N^b_\beta
+{}^{\Im}N^a_\alpha{}^{\Im}N^b_\beta]&\Delta^{-1}T^{-1}_{ab}{}^{\Re}N^b_\alpha&\Delta^{-1}T^{-1}_{ab}{}^{\Im}N^b_\alpha\\
\Delta^{-1}T^{-1}_{ab}{}^{\Re}N^b_\beta&\Delta^{-1}T^{-1}_{ab}&0\\
\Delta^{-1}T^{-1}_{ab}{}^{\Im}N^b_\beta&0&\Delta^{-1}T^{-1}_{ab}\end{array}\right)\end{equation}
and $A(\alpha,a,a), B(\beta,b,b)=1,\cdots,D+1(=d+2n)$. Obviously,
they are equivalent to
\begin{equation}ds^2=h_{\alpha\beta}dx^\alpha\,dx^\beta
+\Delta^{-1}[dz^a+(\mathbf{t}\cdot\mathcal{A})^{ac}z_c]^\dag\,T^{-1}_{ab}[dz^b+(\mathbf{t}\cdot\mathcal{A})^{bd}z_d]\end{equation}
with constraint
\begin{equation}{z^a}^\dag\,z_a=1.\end{equation}
In above IIB supergravity ansatz it seems that an SU(3) isometric
group ansatz can be embeded in it.
\subsection{Kaluza-Klein monopole and instanton}
Because that the Gauss-Codazzi-Ricci equations depend on gauge
potential $\mathcal{A}$, we have pointed that these equations may
be set up in distinct neighborhoods.\cite{Wang} For 11-dimensional
Kaluza-Klein monopole the metric is denoted as
\begin{equation}ds^2_{11}=e^{-\frac{\phi}{6}}ds^2_{10}+e^{\frac{4\phi}{3}}(dx^{10}+\mathcal{A}^\pm)^2,
\end{equation} and
\begin{equation}ds^2_{10}=e^{\frac{\phi}{6}}dx^\mu\,dx_\mu+e^{-\frac{7\phi}{6}}ds^2_3,\qquad\mu=0,\cdots,6\end{equation}
\begin{equation}ds^2_3=dy_idy_i=dr^2+r^2d\theta^2+r^2sin^2\theta\,d{\varphi}^2\\
  \qquad\,r=\sqrt{y_iy_i},\quad\,i=1,2,3\end{equation}
in which $\phi$ is a dilaton and gauge fields will write in the
Wu-Yang gauge\cite{Wu}
\begin{equation}\mathcal{A}^\pm=\frac{Q_m}{2r(y_3\pm\,r)}(y_1dy_2-y_2dy_1)\\
=\frac{1}{2}Q_m(\pm-cos\theta)d{\varphi},\end{equation}
\begin{equation}\mathcal{F}_2=\frac{1}{\sqrt{-h}}\frac{Q_m}{r^2}*(d^7x\wedge\,dr)\\
=\frac{Q_m}{2r^3}\epsilon_{ijk}y_idy_jdy_k=d\mathcal{A}^\pm.\end{equation}
These construct a monopole bundle over base space$S^2$ with fiber
U(1).\\As for instanton we have to look for a fiber bundle over
base space $S^4$ with fiber SU(2). Starting from eq.(75) we set
$\mathbf{t}=\frac{1}{2i}\mathbf{\tau}$ then we need to take a
$S^4$ part out of the d-dimensional metric . It would be better to
choose the polar coordinates of 5-dimensional de Sitter space with
radius $a/2$\cite{Eguchi}
\begin{equation}ds^2_4=[dr^2+r^2(\sigma^2_1+\sigma^2_2+\sigma^2_3)]/(1+\frac{r^2}{a^2})^2\end{equation}
in which
\begin{equation}\sigma_i=\frac{1}{r^2}(x_idx_0-x_0dx_i+\epsilon_{ijk}x_jdx_k),
\qquad r^2=x^\mu\,x_\mu\quad\mu=0,1,2,3.\end{equation} The gauge
potential of BPST instanton will be
\begin{equation}\frac{1}{2i}\mathbf{\tau}\cdot\mathcal{A}^{(+)}_\mu=\frac{r^2}{r^2+a^2}i\mathbf{\sigma}\cdot\mathbf{\tau}
=-i\frac{\sigma_{\mu\nu}x^\nu}{r^2+a^2},\end{equation} in "north"
hemisphere of $S^4$; and \begin{equation}
\frac{1}{2i}\mathbf{\tau}\cdot\mathcal{A}^{(-)}_\mu=\frac{a^2}{r^2+a^2}i\mathbf{\sigma}\cdot\mathbf{\tau}
=-i\frac{a^2\bar{\sigma}_{\mu\nu}x^\nu}{r^2(r^2+a^2)}\end{equation}
in "south" hemisphere of $S^4$, where
\begin{equation}\sigma_{ij}=\bar{\sigma}_{ij}=\frac{1}{2}\epsilon_{ijk}\tau_k,\quad
\sigma_{0i}=-\bar{\sigma}_{0i}=\frac{1}{2}\tau_i,\quad\sigma_{\mu\nu}=-\sigma_{\nu\mu}.\end{equation}
The field strengthes are
\begin{equation}\mathcal{F}^{(+)}=\frac{2ia^2\tau_k}{(r^2+a^2)^2}(dr\wedge\,r\sigma_k+\frac{1}{2}r^2\epsilon_{kij}\sigma_i\wedge\sigma_j),\end{equation}
\begin{equation}\mathcal{F}^{(-)}=h\mathcal{F}^{(+)}h^{-1},\qquad\,h=\frac{t-ix\cdot\tau}{r}.\end{equation}
Of course, we can also use the t'Hooft or Jackiw-Nohl-Rebbi
multiple instanton solution or other instanton solution.

\end{document}